# Single crystal diamond membranes containing germanium vacancy color centers

Kerem Bray[1], Blake Regan[1], Aleksandra Trycz[1], Rodolfo Previdi[1], Gediminas Seniutinas[2], Kumaravelu Ganesan[3], Mehran Kianinia[1], Sejeong Kim[1], and Igor Aharonovich[1*]

[1.] School of Mathematical and Physical Sciences, University of Technology Sydney, Ultimo, NSW, 2007, Australia

[2.] Paul Scherrer Institut, Villigen PSI, 5232, Switzerland

[3.] School of Physics, University of Melbourne, Vic, 3010, Australia

*corresponding author igor.aharonovich@uts.edu.au

**Abstract**

Single crystal diamond membranes that host optically active emitters are highly attractive components for integrated quantum nanophotonics. In this work we demonstrate bottom-up synthesis of single crystal diamond membranes containing the germanium vacancy (GeV) color centers. We employ a lift-off technique to generate the membranes and perform chemical vapour deposition in a presence of germanium oxide to realize the in-situ doping. Finally, we show that these membranes are suitable for engineering of photonic resonators such as microring cavities with quality factors of ~ 1500. The robust and scalable approach to engineer single crystal diamond membranes containing emerging color centers is a promising pathway for realization of diamond integrated quantum nanophotonic circuits on a chip.

**Main text**

Color centers in diamond are attractive building blocks for solid state integrated quantum photonics[1-2]. So far, significant effort has been devoted to study the nitrogen vacancy (NV) center due to its long electron spin coherence time and hence its applicability as a spin qubit and a sensor[3-4]. However, its optical properties are not ideal as it has a large phonon sideband with only ~ 4% of the fluorescence being emitted into the zero phonon line (ZPL). The NV center also has a permanent dipole moment which makes it highly susceptible to external electromagnetic fluctuations. These fluctuations result in ultrafast spectral diffusion of its optical transitions, and hinder the generation rate of indistinguishable photons[5-6].

To overcome these shortfalls, new color centers with inversion symmetry based on group IV elements in diamond has been recently explored. Following on detailed investigations of the negatively charged silicon vacancy (SiV)[7-14], other color centers including germanium vacancy (GeV)[15-20] and more recently tin vacancy (SnV)[21-22] have been identified. The defects have an inversion symmetry and therefore are less vulnerable to strain and electromagnetic fluctuations. Indeed, recent reports showed that generation of Fourier Transform limited photons is possible employing SiV centers in both nanodiamonds and bulk, and fabricated nanostructures using a reactive ion etching process[12, 23-25]. On the other hand, the advantage of the GeV and the SnV is their larger energy splitting in the ground state that reduces the phonon mediated spin mixing, and perform coherent spin manipulation at liquid Helium temperatures[26-27]. In addition, the GeV color center is believed to have a much higher quantum efficiency compared to the SiV, therefore being advantageous for applications in quantum optics and nanophotonics. To employ these color centers in integrated nanophotonic devices, an interface between the spin and the emitted photons is needed. Towards this goal, a promising avenue is an integration of these color centers with photonic resonators[1, 11, 28-31].

In this work, we demonstrate a robust method to engineer microdisk and microring resonators containing the GeV color centers. The resonators are fabricated from single crystal diamond membranes (~300 nm) that were overgrown in a presence of germanium to achieve pristine diamond crystal that contains homogeneous ensemble of bright GeV emitters.

The diamond membranes ([100] faceted) were generated using ion implantation and liftoff, as described in details previously[32-34]. To generate the GeV color centers, the membranes were transferred onto a sapphire substrate using a liquid droplet of germanium dioxide ($GeO_2$) powder (SigmaAldrich) – as shown in figure 1a. The droplet contains a 0.5 mg/mL solution of $GeO_2$ powder in ethanol. The sapphire substrate with the $GeO_2$ covered membrane was then placed in a microwave plasma chemical vapor deposition (MPCVD) chamber, along with a ~1 $mm^2$ piece of metallic germanium positioned approximately ~ 1 cm away from the diamond membrane (figure 1b). We found that when the metallic germanium is the only source during growth, strong GeV color centres appeared preferentially along the edges of the membrane. On the other hand, the addition of the $GeO_2$ increases the incorporation of Ge into the surface of the growing diamond. The CVD growth conditions were a hydrogen/methane ratio of 100:1 at 60 Torr, a microwave power of 900W for 10 minutes to fabricate a ~ 400 nm intrinsic diamond layer that contains homogeneous ensemble of GeV color centers. The incorporation of the germanium into the diamond occurs via the plasma phase. Upon the microwave plasma, the oxide and the metallic source melt, producing germanium vapour that is then homogeniously incorporated into the growing diamond. To remove the original diamond that did not contain the GeV color centers, the membranes were flipped and thinned by Inductively Coupled Plasma - Reactive Ion Etching (ICP-RIE) in the presence of oxygen and $SF_6$ at a pressure of 45 mTorr for 18 minutes (figure 1c). This fabrication process results in a single crystal diamond membrane with a thickness of 300 nm (figure 1d).

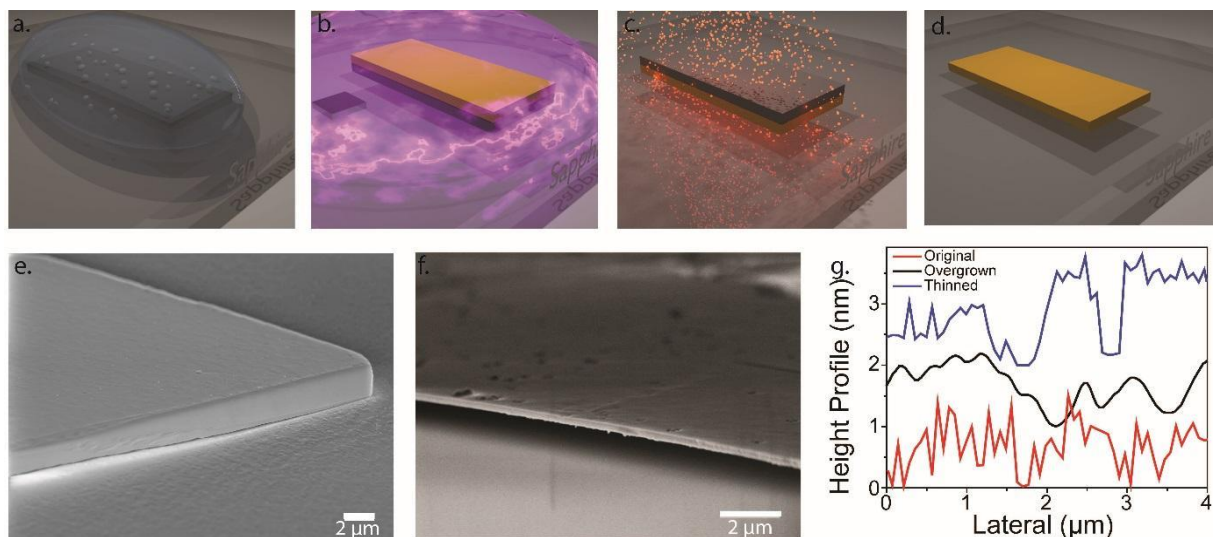

*Figure 1* *Schematic illustration of the fabrication process to engineer single crystal diamond membranes containing GeV color centers. (a) An electrochemically etched single crystal diamond membrane (dark grey) is transferred to a sapphire substrate using a 0.5 mg/mL $GeO_2$ solution in ethanol. (b) Subsequent MPCVD overgrowth with an additional metallic germanium source nearby. (c) The sample is flipped and thinned by ICP-RIE, to remove the*

*original material to generate (d) pristine diamond membrane with GeV centers (yellow). (e, f) Representative SEM images of the overgrown and the final (thinned) diamond membrane, respectively. The thicknesses of the original and the thinned membrane are ~ 2.10 μm and ~ 0.3 μm, respectively. (g) AFM scans of the surface roughness of the original (red curve), overgrown (black curve) and thinned (dark blue curve) diamond membranes. The surface roughness of these membranes is ~1 nm. The curves are off-set in 1 nm intervals for clarity.*

Representative side view scanning electron microscope (SEM) images of the overgrown and the final (thinned) membrane are shown in figure 1e and figure 1f, respectively. The surface roughness of the original, overgrown and thinned membrane was determined by an atomic force microscope (AFM), after each fabrication step. The slight curve observed in figure 1f is likely due to residual strain within the diamond membrane. The AFM measured surface profiles are shown in figure 1g, and reveal minimal surface roughness of ~ 1 nm. Therefore, the overgrowth method in the presence of germanium does not hinder the smoothness nor reduces the quality of the final single crystal diamond membrane.

To study the optical properties of the diamond membranes, photoluminescence (PL) measurements were recorded under a continuous-wave 532 nm laser (Gem 532, Laser Quantum) excitation at room and cryogenic temperatures using a standard confocal microscope. Figure 2a shows a RT PL spectrum of the thin (~ 300 nm) diamond membrane with the strong GeV ZPL at 602 nm and a full width at half maximum (FWHM of ~ 0.8 nm). Inset, is the schematic illustration of the GeV defect, that consists of an interstitial Ge atom splitting two vacancies in a diamond lattice. In some locations along the membrane, an additional peak around ~ 640 nm was also observed. This luminescence signal can potentially be attributed to the neutral charge state of the GeV emitters. This is plausible as the overgrowth occurs under hydrogen presence that is known to introduce band bending and cause increased emission from the neutrally charged NV centers[35-36]. Furthermore, the ~ 640 nm peak was not observed in overgrown membranes without a germanium source, and likewise, this peak was never observed without the ZPL at 602 nm. However, at this point we do not have a concrete evidence for the unambiguous identification of the GeV charge state.

The PL spectrum from the same membrane recorded at 15 K is shown in figure 2b. At cryogenic temperature, the PL spectrum from the GeV emitters in low strain material splits into four lines corresponding to the transitions from the low (high) branch of the ground state to the low (high) branch of the excited state[15, 19]. The four distinct PL lines are clearly visible in figure 2b, corresponding to the optical transitions of the GeV ZPL. Inset, is the schematic of the electronic level structure of the GeV, with the splitting in the ground and excited states. Based on our measurements, the ground state splitting is $\Delta E_{1 \rightarrow 2}$ ~ 165 GHz while the excited state splitting is $\Delta E_{3 \rightarrow 4}$ ~ 1159 GHz in accord with the reported literature[15, 19].

An important prerequisite for scalable devices is a homogeneous distribution of the color centers. This is important for various applications such as ensemble magnetometry or cavity cooperativity schemes[37-38], as well as for practical aims to enable patterning of multiple devices on a single membrane. Figure 3c shows a confocal scan of the diamond membrane containing the GeV emitters. The RT spectra in figure 3d were recorded at ~ 10 μm intervals across the membrane, from the locations marked with numbers 1 – 6. Each point shows the typical GeV peak at ~ 602 nm (figure 3d), confirming a homogeneous distribution of the emitters.

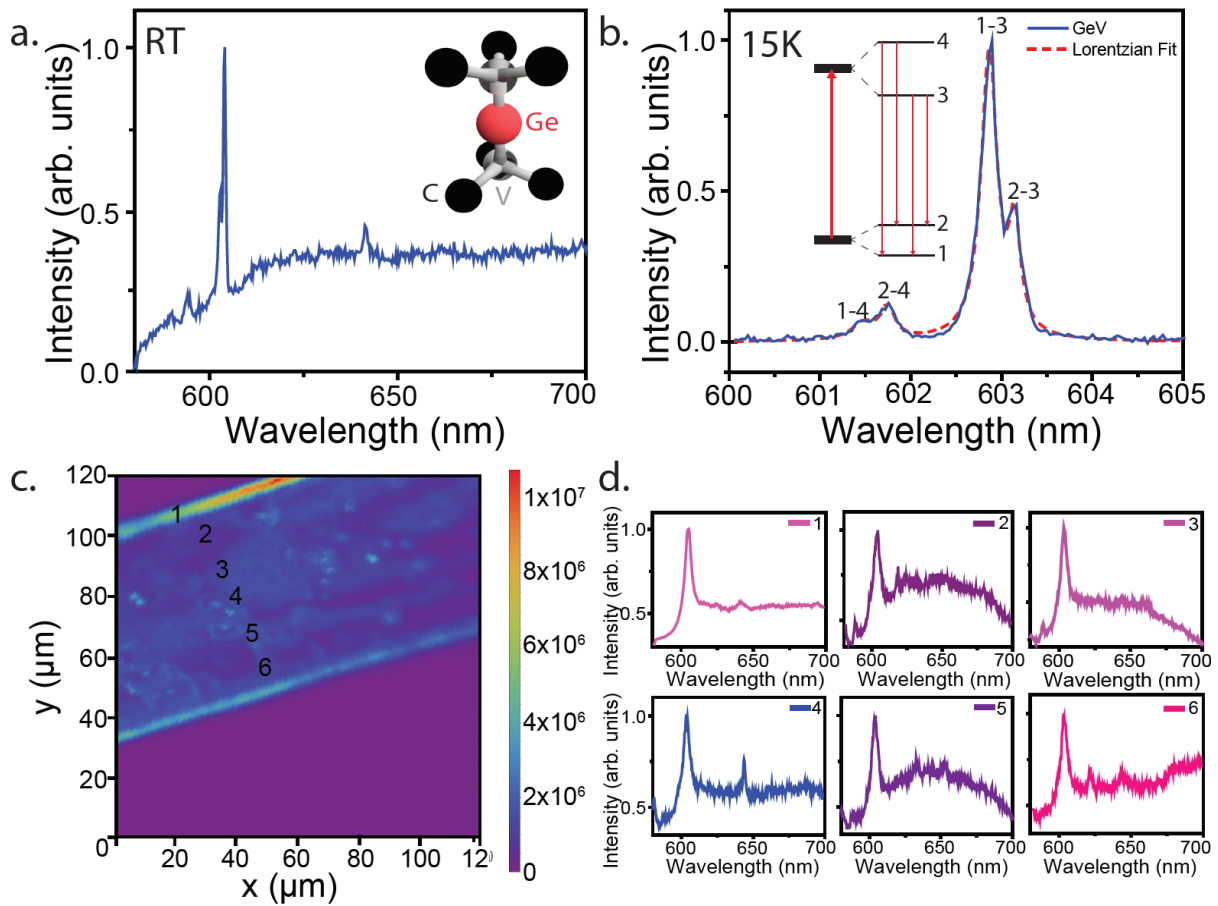

*Figure 2. Spectra of the diamond membranes doped with GeV. (a) A RT spectrum showing a characteristic peak at 602 nm. Inset, schematic structure of the GeV color center. (b) A spectrum of the GeV taken at 15K (blue curve), with the characteristic splitting corresponding to the transitions from the excited to the ground state manifolds. The red dashed line is the Lorentzian fit. Inset, is the schematic of the electronic level structure of the GeV. (c) A confocal map of the overgrown diamond membrane where each number corresponds to a spectrum in (d) showing the homogeneous distribution of the GeV emitters. (d) Six representative RT spectra of the GeV emitters in the diamond membrane, showing a homogeneous distribution of the GeV emission.*

Finally, we utilize the diamond membranes to fabricate microdisk and micro-ring cavities which are commonly used as photonic resonators[39]. A hydrogen silsesquioxane (HSQ) hard mask was deposited on top of the overgrown and thinned membrane and an array of microdisk cavities were patterned using Electron Beam Lithography (EBL). Figure 3a shows a low magnification SEM image of the patterned devices. The microdisks have a 2.5 μm radius and 5 μm spacing (edge to edge) between each disk (figure 3a), while the micro-rings have a 5 μm outer diameter and 400 nm inner diameter, respectively. The diamond membrane is outlined for clarity with a blue dashed line. Figure 3b, c show a false color SEM image of the diamond microdisk and a diamond microring, respectively, with a thickness of ~ 300 nm, resting on the SiO$_2$ substrate.

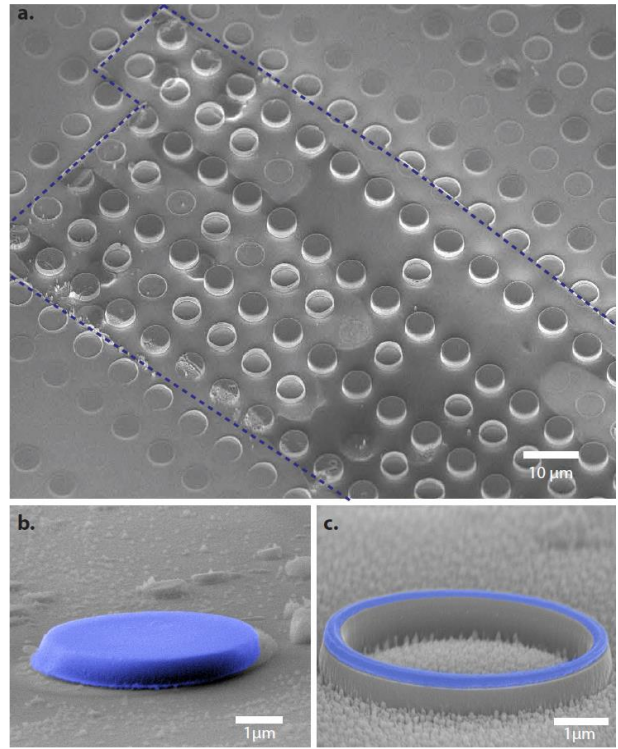

*Figure 3. Diamond microdisk and microring cavities fabricated from the overgrown diamond membranes containing GeV color ceters. (a) Low magnification SEM image of microring and microdisk resonators with a diameter of 5 µm etched into the diamond membrane (blue dashed line). (b, c) False color SEM images of a diamond microdisk and a diamond microring, respectively. The diamond are false colored blue for clarity.*

Figure 4a shows a PL spectrum recorded from one of the diamond microrings. Both the microdisks and the microrings support propagation of whispering gallery modes (WGMs) that are manifested as periodic peaks within the spectrum. For the current microring size, the spacing between adjacent modes should be ~ 14 nm. The PL emission is decorated by narrow lines, which are the modes of the optical cavity, while a strong GeV ZPL is visible at 602 nm. Quality factors *Q* as high as 1500 were observed from these devices. A particular mode that is in a close proximity to the GeV ZPL at 596 nm exhibits a *Q* ~ 750. A high resolution spectrum from two modes at ~ 750 nm (figure 4b) show *Q* ~ 1500. The losses occur due to an increased roughness from the imperfect HSQ mask removal and from the leakage to the underlying SiO$_2$ substrate.

Figure 4c shows the modelling result of the diamond microring (*n*=2.4) of the same geometry resting on a silicon dioxide substrate (*n*=1.5) using a finite-difference time-domain (FDTD) method (Lumerical inc.). The structure support 1$^{st}$ order radial Transverse electric (TE) modes as indicated in the spectrum. As an example, electric field profile of TE$_{1,74}$ mode is shown in inset. Other peaks in the spectrum are higher order radial modes including 2$^{nd}$ and 3$^{rd}$ radial modes. TE$_{2,62}$ is shown in the inset. A very good matching between the measured the simulated spectra is observed. The dominant presence of the 1$^{st}$ order modes at the longer wavelength range while the lower wavelength range also supports higher order modes as evidenced by closed spacing of the peaks.

The fabricated devices enable higher collection of the emitted photons from the GeV color center. Figure 4d shows an emission enhancement from the membrane as compared to the fabricated microring cavities. An enhancement of ~ 14 times is clearly observed, after measuring more than ten different devices. Such a brightness will be advantageous for employing the GeV color centers in practical nanophotonics devices.

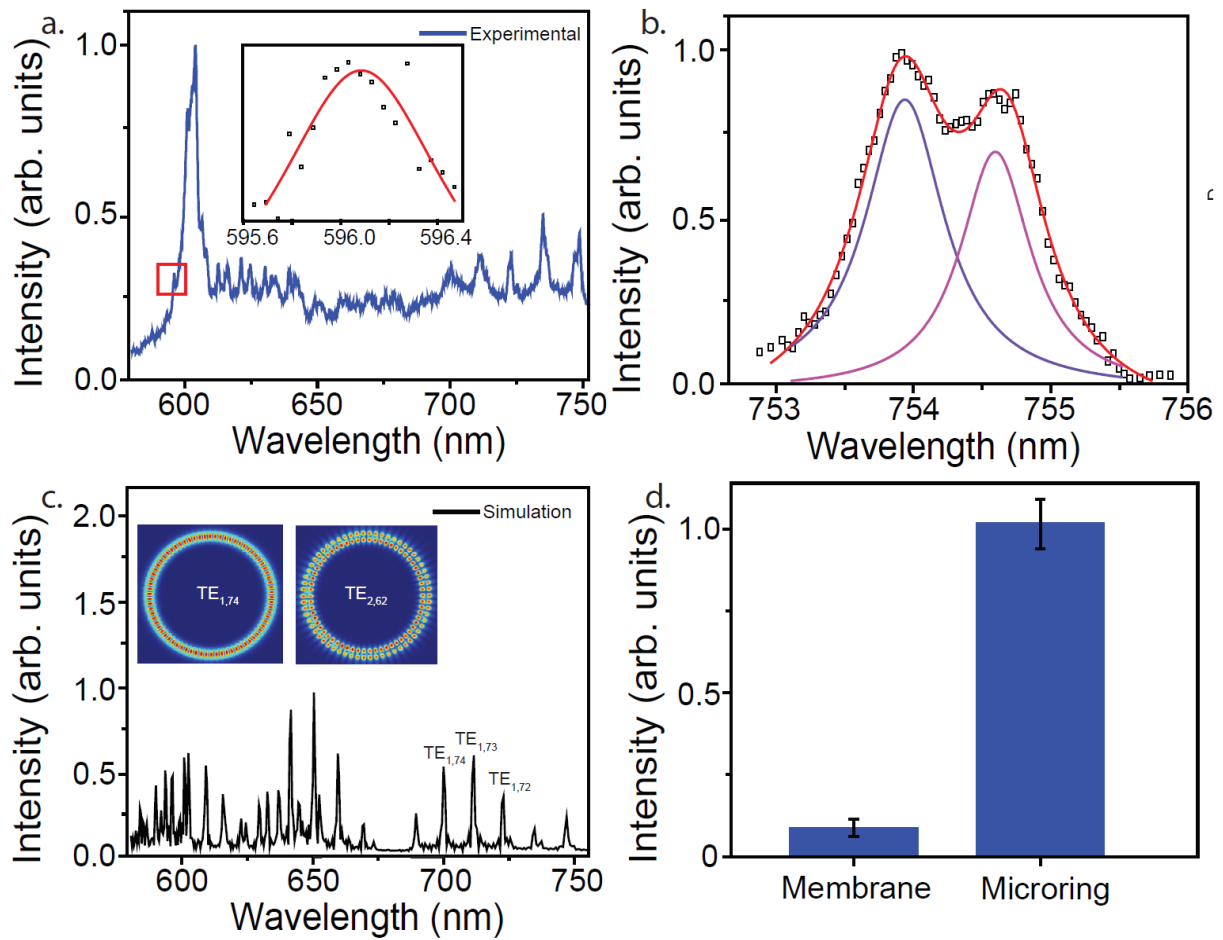

*Figure 4. (a) PL spectrum recorded from the diamond microring cavity under 532 nm excitation at room temperature. WGMs are clearly visible. The ZPL of the GeV center is observed at 602 nm. Inset, a zoom in spectrum of the mode at 596 nm. The red line is a Lorentzian fit. (b) a spectrum of a WGM at ~ 755 nm, showing a Q ~ 1500. (b) Simulation data of the microring sample. Inset, FDTD field profile of the 1$^{st}$ (left) and 2$^{nd}$ (right) order mode. (d) comparison of recorded emission from GeV color center recorded from the membrane and from a micro-ring cavity, showing an enhancement of ~ 14 times.*

In conclusion, we have described a robust method to engineer optically-active, high-quality single crystal diamond membranes containing a homogeneous ensemble of GeV emitters. We further employed these membranes to produce microdisk and microring cavities containing GeV emitters with $Q$ values of ~ 1500. Note that the presented method of doping the diamond membranes is advantageous and can be applied also for bulk crystals or nanodiamonds, without degrading the overgrown crystal quality. Furthermore, similar technique can be explored to achieve doping of diamond membranes with other emerging colour centres such as tin related emitters[21-22] and other narrow band emitters[40]. Our results will accelerate the integration of new colour centres in diamond with scalable photonic devices to achieve on-chip quantum nanophotonic circuitry.

**Acknowledgements**
Financial support from the Australian Research Council (via DP180100077), the Asian Office of Aerospace Research and Development grant (FA2386-17-1-4064) and Office of Naval Research Global (N62909-18-1-2025). I.A. gratefully acknowledges JSPS Invitation Fellowships [S16712]. This research is supported by an Australian Government Research Training Program Scholarship. G. S. acknowledges funding from the EU-H2020 Research and Innovation program under Grant Agreement No. 654360 NFFA-Europe.

**References**
1. Sipahigil, A.; Evans, R. E.; Sukachev, D. D.; Burek, M. J.; Borregaard, J.; Bhaskar, M. K.; Nguyen, C. T.; Pacheco, J. L.; Atikian, H. A.; Meuwly, C.; Camacho, R. M.; Jelezko, F.; Bielejec, E.; Park, H.; Lončar, M.; Lukin, M. D., An Integrated Diamond Nanophotonics Platform for Quantum Optical Networks. *Science* **2016**.
2. Kalb, N.; Reiserer, A. A.; Humphreys, P. C.; Bakermans, J. J. W.; Kamerling, S. J.; Nickerson, N. H.; Benjamin, S. C.; Twitchen, D. J.; Markham, M.; Hanson, R., Entanglement Distillation between Solid-State Quantum Network Nodes. *Science* **2017,** *356*, 928-932.
3. Childress, L.; Hanson, R., Diamond Nv Centers for Quantum Computing and Quantum Networks. *MRS Bulletin* **2013,** *38*, 134-138.
4. Aslam, N.; Pfender, M.; Neumann, P.; Reuter, R.; Zappe, A.; Fávaro de Oliveira, F.; Denisenko, A.; Sumiya, H.; Onoda, S.; Isoya, J.; Wrachtrup, J., Nanoscale Nuclear Magnetic Resonance with Chemical Resolution. *Science* **2017,** *357*, 67-71.
5. Acosta, V. M.; Santori, C.; Faraon, A.; Huang, Z.; Fu, K. M. C.; Stacey, A.; Simpson, D. A.; Ganesan, K.; Tomljenovic-Hanic, S.; Greentree, A. D.; Prawer, S.; Beausoleil, R. G., Dynamic Stabilization of the Optical Resonances of Single Nitrogen-Vacancy Centers in Diamond. *Phys. Rev. Lett.* **2012,** *108*.
6. Schmidgall, E. R.; Chakravarthi, S.; Gould, M.; Christen, I. R.; Hestroffer, K.; Hatami, F.; Fu, K.-M. C., Frequency Control of Single Quantum Emitters in Integrated Photonic Circuits. *Nano Lett.* **2018**.
7. Neu, E.; Steinmetz, D.; Riedrich-Moeller, J.; Gsell, S.; Fischer, M.; Schreck, M.; Becher, C., Single Photon Emission from Silicon-Vacancy Centres in Cvd-Nano-Diamonds on Iridium *New J. Phys.* **2011,** *13*, 025012.
8. Becker, J. N.; Görlitz, J.; Arend, C.; Markham, M.; Becher, C., Ultrafast All-Optical Coherent Control of Single Silicon Vacancy Colour Centres in Diamond. *Nat. Commun.* **2016,** *7*, 13512.
9. Hepp, C.; Müller, T.; Waselowski, V.; Becker, J. N.; Pingault, B.; Sternschulte, H.; Steinmüller-Nethl, D.; Gali, A.; Maze, J. R.; Atatüre, M.; Becher, C., Electronic Structure of the Silicon Vacancy Color Center in Diamond. *Phys. Rev. Lett.* **2014,** *112*, 036405.
10. Zhou, Y.; Rasmita, A.; Li, K.; Xiong, Q.; Aharonovich, I.; Gao, W.-b., Coherent Control of a Strongly Driven Silicon Vacancy Optical Transition in Diamond. *Nat. Commun.* **2017,** *8*, 14451.
11. Zhang, J. L.; Ishiwata, H.; Babinec, T. M.; Radulaski, M.; Müller, K.; Lagoudakis, K. G.; Dory, C.; Dahl, J.; Edgington, R.; Soulière, V.; Ferro, G.; Fokin, A. A.; Schreiner, P. R.; Shen, Z.-X.; Melosh, N. A.; Vučković, J., Hybrid Group Iv Nanophotonic Structures Incorporating Diamond Silicon-Vacancy Color Centers. *Nano Lett.* **2016,** *16*, 212-217.
12. Rogers, L. J.; Jahnke, K. D.; Teraji, T.; Marseglia, L.; Müller, C.; Naydenov, B.; Schauffert, H.; Kranz, C.; Isoya, J.; McGuinness, L. P.; Jelezko, F., Multiple Intrinsically Identical Single-Photon Emitters in the Solid State. *Nat. Commun.* **2014,** *5*, 4739.

13. Arend, C.; Becker, J. N.; Sternschulte, H.; Steinmüller-Nethl, D.; Becher, C., Photoluminescence Excitation and Spectral Hole Burning Spectroscopy of Silicon Vacancy Centers in Diamond. *Phys. Rev. B* **2016,** *94*, 045203.
14. Kay, D. J.; Alp, S.; Jan, M. B.; Marcus, W. D.; Mathias, M.; Lachlan, J. R.; Neil, B. M.; Mikhail, D. L.; Fedor, J., Electron–Phonon Processes of the Silicon-Vacancy Centre in Diamond. *New J. Phys.* **2015,** *17*, 043011.
15. Bhaskar, M. K.; Sukachev, D. D.; Sipahigil, A.; Evans, R. E.; Burek, M. J.; Nguyen, C. T.; Rogers, L. J.; Siyushev, P.; Metsch, M. H.; Park, H.; Jelezko, F.; Lončar, M.; Lukin, M. D., Quantum Nonlinear Optics with a Germanium-Vacancy Color Center in a Nanoscale Diamond Waveguide. *Phys. Rev. Lett.* **2017,** *118*, 223603.
16. Siyushev, P.; Metsch, M. H.; Ijaz, A.; Binder, J. M.; Bhaskar, M. K.; Sukachev, D. D.; Sipahigil, A.; Evans, R. E.; Nguyen, C. T.; Lukin, M. D.; Hemmer, P. R.; Palyanov, Y. N.; Kupriyanov, I. N.; Borzdov, Y. M.; Rogers, L. J.; Jelezko, F., Optical and Microwave Control of Germanium-Vacancy Center Spins in Diamond. *Phys. Rev. B* **2017,** *96*, 081201.
17. Boldyrev, K. N.; Mavrin, B. N.; Sherin, P. S.; Popova, M. N., Bright Luminescence of Diamonds with Ge-V Centers. *J. Lumin.* **2018,** *193*, 119-124.
18. Palyanov, Y. N.; Kupriyanov, I. N.; Borzdov, Y. M.; Khokhryakov, A. F.; Surovtsev, N. V., High-Pressure Synthesis and Characterization of Ge-Doped Single Crystal Diamond. *Crystal Growth & Design* **2016,** *16*, 3510-3518.
19. Iwasaki, T.; Ishibashi, F.; Miyamoto, Y.; Doi, Y.; Kobayashi, S.; Miyazaki, T.; Tahara, K.; Jahnke, K. D.; Rogers, L. J.; Naydenov, B.; Jelezko, F.; Yamasaki, S.; Nagamachi, S.; Inubushi, T.; Mizuochi, N.; Hatano, M., Germanium-Vacancy Single Color Centers in Diamond. *Scientific Reports* **2015,** *5*, 12882.
20. Ralchenko, V. G.; Sedov, V. S.; Khomich, A. A.; Krivobok, V. S.; Nikolaev, S. N.; Savin, S. S.; Vlasov, I. I.; Konov, V. I., Observation of the Ge-Vacancy Color Center in Microcrystalline Diamond Films. *Bulletin of the Lebedev Physics Institute* **2015,** *42*, 165-168.
21. Tchernij, S. D.; Herzig, T.; Forneris, J.; Küpper, J.; Pezzagna, S.; Traina, P.; Moreva, E.; Degiovanni, I. P.; Brida, G.; Skukan, N.; Genovese, M.; Jakšić, M.; Meijer, J.; Olivero, P., Single-Photon-Emitting Optical Centers in Diamond Fabricated Upon Sn Implantation. *ACS Photonics* **2017,** *4*, 2580-2586.
22. Iwasaki, T.; Miyamoto, Y.; Taniguchi, T.; Siyushev, P.; Metsch, M. H.; Jelezko, F.; Hatano, M., Tin-Vacancy Quantum Emitters in Diamond. *Phys. Rev. Lett.* **2017,** *119*, 253601.
23. Li, K.; Zhou, Y.; Rasmita, A.; Aharonovich, I.; Gao, W. B., Nonblinking Emitters with Nearly Lifetime-Limited Linewidths in Cvd Nanodiamonds. *Physical Review Applied* **2016,** *6*, 024010.
24. Evans, R. E.; Sipahigil, A.; Sukachev, D. D.; Zibrov, A. S.; Lukin, M. D., Narrow-Linewidth Homogeneous Optical Emitters in Diamond Nanostructures Via Silicon Ion Implantation. *Physical Review Applied* **2016,** *5*, 044010.
25. Rogers, L. J.; Wang, O.; Liu, Y.; Antoniuk, L.; Osterkamp, C.; Davydov, V. A.; Agafonov, V. N.; Filipovski, A. B.; Jelezko, F.; Kubanek, A., Single Siv− Centers in Low-Strain Nanodiamonds with Bulk-Like Spectral Properties and Nano-Manipulation Capabilities. *https://arxiv.org/pdf/1802.03588* **2018**.
26. Becker, J. N.; Pingault, B.; Groß, D.; Gündoğan, M.; Kukharchyk, N.; Markham, M.; Edmonds, A.; Atatüre, M.; Bushev, P.; Becher, C., All-Optical Control of the Silicon-Vacancy Spin in Diamond at Millikelvin Temperatures. *Phys. Rev. Lett.* **2018,** *120*, 053603.
27. Sukachev, D. D.; Sipahigil, A.; Nguyen, C. T.; Bhaskar, M. K.; Evans, R. E.; Jelezko, F.; Lukin, M. D., Silicon-Vacancy Spin Qubit in Diamond: A Quantum Memory Exceeding 10 Ms with Single-Shot State Readout. *Phys. Rev. Lett.* **2017,** *119*, 223602.
28. Schröder, T.; Mouradian, S. L.; Zheng, J.; Trusheim, M. E.; Walsh, M.; Chen, E. H.; Li, L.; Bayn, I.; Englund, D., Quantum Nanophotonics in Diamond [Invited]. *J. Opt. Soc. Am. B* **2016,** *33*, B65-B83.
29. Mouradian, S. L.; Schröder, T.; Poitras, C. B.; Li, L.; Goldstein, J.; Chen, E. H.; Walsh, M.; Cardenas, J.; Markham, M. L.; Twitchen, D. J.; Lipson, M.; Englund, D., Scalable Integration of Long-Lived Quantum Memories into a Photonic Circuit. *Physical Review X* **2015,** *5*, 031009.
30. Burek, M. J.; Meuwly, C.; Evans, R. E.; Bhaskar, M. K.; Sipahigil, A.; Meesala, S.;

Machielse, B.; Sukachev, D. D.; Nguyen, C. T.; Pacheco, J. L.; Bielejec, E.; Lukin, M. D.; Lončar, M., Fiber-Coupled Diamond Quantum Nanophotonic Interface. *Physical Review Applied* **2017,** *8*, 024026.
31. Riedrich-Möller, J.; Arend, C.; Pauly, C.; Mücklich, F.; Fischer, M.; Gsell, S.; Schreck, M.; Becher, C., Deterministic Coupling of a Single Silicon-Vacancy Color Center to a Photonic Crystal Cavity in Diamond. *Nano Lett.* **2014,** *14*, 5281-5287.
32. Magyar, A.; Lee, J. C.; Limarga, A. M.; Aharonovich, I.; Rol, F.; Clarke, D. R.; Huang, M. B.; Hu, E. L., Fabrication of Thin, Luminescent, Single-Crystal Diamond Membranes. *Appl. Phys. Lett.* **2011,** *99*, 081913.
33. Lee, J. C.; Aharonovich, I.; Magyar, A. P.; Rol, F.; Hu, E. L., Coupling of Silicon-Vacancy Centers to a Single Crystal Diamond Cavity. *Opt. Express* **2012,** *20*, 8891-8897.
34. Olivero, P.; Rubanov, S.; Reichart, P.; Gibson, B. C.; Huntington, S. T.; Rabeau, J.; Greentree, A. D.; Salzman, J.; Moore, D.; Jamieson, D. N.; Prawer, S., Ion-Beam-Assisted Lift-Off Technique for Three-Dimensional Micromachining of Freestanding Single-Crystal Diamond. *Adv. Mater.* **2005,** *17*, 2427.
35. Stacey, A.; Karle, T. J.; McGuinness, L. P.; Gibson, B. C.; Ganesan, K.; Tomljenovic-Hanic, S.; Greentree, A. D.; Hoffman, A.; Beausoleil, R. G.; Prawer, S., Depletion of Nitrogen-Vacancy Color Centers in Diamond Via Hydrogen Passivation. *Appl. Phys. Lett.* **2012,** *100*, -.
36. Fu, K. M. C.; Santori, C.; Barclay, P. E.; Beausoleil, R. G., Conversion of Neutral Nitrogen-Vacancy Centers to Negatively Charged Nitrogen-Vacancy Centers through Selective Oxidation. *Appl. Phys. Lett.* **2010,** *96*.
37. Zhu, X.; Saito, S.; Kemp, A.; Kakuyanagi, K.; Karimoto, S.-i.; Nakano, H.; Munro, W. J.; Tokura, Y.; Everitt, M. S.; Nemoto, K.; Kasu, M.; Mizuochi, N.; Semba, K., Coherent Coupling of a Superconducting Flux Qubit to an Electron Spin Ensemble in Diamond. *Nature* **2011,** *478*, 221-224.
38. Zhong, T.; Kindem, J. M.; Miyazono, E.; Faraon, A., Nanophotonic Coherent Light–Matter Interfaces Based on Rare-Earth-Doped Crystals. *Nature Communications* **2015,** *6*, 8206.
39. Faraon, A.; Santori, C.; Huang, Z. H.; Acosta, V. M.; Beausoleil, R. G., Coupling of Nitrogen-Vacancy Centers to Photonic Crystal Cavities in Monocrystalline Diamond. *Phys. Rev. Lett.* **2012,** *109*.
40. Bray, K.; Sandstrom, R.; Elbadawi, C.; Fischer, M.; Schreck, M.; Shimoni, O.; Lobo, C.; Toth, M.; Aharonovich, I., Localization of Narrowband Single Photon Emitters in Nanodiamonds. *ACS Applied Materials & Interfaces* **2016,** *8*, 7590-7594.